# A Stochastic Analysis of Steady and Transient Heat Conduction in Random Media Using a Homogenization Approach

Zhijie Xu[1,a]

1. Energy Resource Recovery & Management, Idaho National Laboratory, Idaho Falls, Idaho 83415 USA. Now at

Computational Mathematics Group, Fundamental and Computational Sciences Directorate, Pacific Northwest National Laboratory, Richland, WA 99352, USA

We present a new stochastic analysis for steady and transient one-dimensional heat conduction problem based on the homogenization approach. Thermal conductivity is assumed to be a random field $K$ consisting of random variables of a total number $N$. Both steady and transient solutions $T$ are expressed in terms of the homogenized solution $\tilde{T}$ and its spatial derivatives $T(x,t) = \tilde{T} + \sum_{n=1}^{\infty} L_n(x) \partial^n \tilde{T} / \partial x^n$, where homogenized solution $\tilde{T}$ is obtained by solving the homogenized equation with effective thermal conductivity. Both mean and variance of stochastic solutions can be obtained analytically for $K$ field consisting of *identically distributed (i.i.d) random* variables. The mean and variance of $T$ are shown to be dependent only on the mean and variance of these i.i.d variables, not the particular form of probability distribution function of i.i.d variables. Variance of temperature field $T$ can be separated into two contributions: the ensemble contribution (through the homogenized temperature $\tilde{T}$); and the configurational contribution (through the random variable $L_n(x)$). The configurational contribution is shown to be proportional to the local gradient of $\tilde{T}$.

a) Electronic mail: zhijie.xu@pnnl.gov, zhijiexu@hotmail.com

Large uncertainty of $T$ field was found at locations with large gradient of $\tilde{T}$ due to the significant configurational contributions at these locations. Numerical simulations were implemented based on a direct Monte Carlo method and good agreement is obtained between numerical Monte Carlo results and the proposed stochastic analysis.

# I. Introduction

The modeling and analysis of heat conduction for any mechanical systems is a fundamental problem that is critical to standard engineering design. Traditional deterministic modeling is based on the assumption that problem input parameters (material properties, boundary and initial conditions, and problem geometry etc.) are known and fully deterministic [1]. However, only using the mean values from the deterministic study for design variables is generally not enough for most engineering practice and a safety factor must be introduced in order to take into account the uncertainty associated with the physical properties, initial, and boundary conditions. This factor of safety provides a measure of the reliability of a particular design. An accurate estimation of the safety factor to be used in the design will significantly reduce the manufacturing and operating cost. Therefore, there is an increasing interest in stochastic analysis and modeling for engineering problems, where uncertainty quantification and sensitivity analysis can be implemented in order to estimate an optimized factor of safety.

As the use of probabilistic design methodology becomes a standard practice for most engineering design, a number of stochastic modeling methods have been developed in different ways. A popular approach is to employ a large number of Monte Carlo simulations (MCS) over a sufficiently large probability space. Statistical information can then be extracted from the outcome of MCS. Since the standard deviation is inversely proportional to the number of samples ($N_{sim}$), a $N_{sim} \approx 500$ is usually required in order to estimate the cumulative distribution function (CDF) [2]. Therefore, this statistical approach often suffers from the enormous computational effort that is required for sufficiently large samples.

Some non-statistical approaches have also been developed to overcome the computational hurdles associated with the direct statistical approach. The Galerkin method applied to stochastic systems has been proposed in [3]. The stochastic version of Taylor expansion, where the stochastic fields are discretized around mean values by a Taylor series expansion, is also known in the literature as the perturbation method [4-6]. As a powerful tool for solving the stochastic partial differential equations (PDEs), the stochastic finite element method (SFEM), an extension of the standard finite element method to the stochastic framework, has been put forward and applied to a number of examples [7-9], where the underlying formulation uses the second-order perturbation for the expansion of the random fields. Another popular approach, called polynomial chaos, is based on the decomposition of the random inputs and solutions into chaos polynomials which do form a complete and orthogonal basis. The original version employs a spectral expansion using Hermite orthogonal polynomials for Gaussian random variables. These approaches have been applied to the uncertainty analysis of heat conduction by a number of authors [10-12]. In contrast to the method mentioned, the new approach is based on the homogenization through expansion of dispersion relation. The original problem is reduced to quantify the uncertainty associated with the ensemble and configurational contributions. Significant computational savings can be achieved, as demonstrated in the numerical examples.

The homogenization approach was applied in our previous work [13, 14] to derive the effective properties and homogenized solutions. Reduction of problem dimensionality can also be achieved through the cross-sectional homogenization using the reduced-boundary-function method [15, 16]. The objective of this paper is to present a stochastic analysis based on the homogenization approach (employing the dispersion relations) presented in [13, 14] to provide some insights of uncertainties associated with steady-state and transient heat conduction, where the thermal conductivity can be any arbitrary random field in principle. The paper is organized as follows. We first present the analysis for steady problem in Section II, followed by the analysis for transient problem using a homogenization approach in Section III. Numerical examples are presented in Section IV.

## II. Stochastic Analysis for Steady-State Heat Conduction

In this section we consider the steady-state heat conduction problem subject to uncertainty in the thermal conductivity. We consider a one-dimensional model problem where the thermal conductivity $K(x)$ is a random field that consists of random variables at different location of $x$. The equation for this problem is written as,

$$\frac{d}{dx}\left[K\frac{dT}{dx}\right] = 0, \qquad x \in [0,1], \tag{1}$$

with Dirichlet boundary conditions

$$T(x=0) = T_0 \text{ and } T(x=1) = T_1, \tag{2}$$

where $x$ is the position and $T$ is the temperature field. The steady-state temperature $T$ is also a stochastic field because of the randomness in thermal conductivity $K$. The objective is to find the statistical properties of $T$ for given random field $K$. From Eq. (1), we can easily find that

$$K\frac{dT}{dx} = A, \tag{3}$$

where $A$ is a constant that can be determined from boundary conditions, where

$$T(x=1) = T_0 + \frac{1}{N}\sum_{n=1}^{N}\frac{dT}{dx} = T_0 + \frac{A}{N}\sum_{n=1}^{N}\frac{1}{K} = T_1. \tag{4}$$

Here $N$ is the total number of discretized random variables in the interval $[0,1]$ after discretization. $K$ is a random field in space. Equation (4) leads to the constant $A$,

$$A = N(T_1 - T_0) \bigg/ \sum_{n=1}^{N} \frac{1}{K}. \tag{5}$$

The temperature field $T$ can be explicitly written as,

$$T(x) = T_0 + \frac{1}{N}\sum_{n=1}^{N_1}\frac{dT}{dx} = T_0 + \frac{A}{N}\sum_{n=1}^{N_1}\frac{1}{K} = T_0 + (T_1-T_0)\sum_{n=1}^{N_1}\frac{1}{K} \bigg/ \sum_{n=1}^{N}\frac{1}{K}, \tag{6}$$

where $N_1 = Nx$ represents the number of random variables between $[0, x]$. By substitution of

$$\sum_{n=1}^{N}\frac{1}{K} = \sum_{n=1}^{N_1}\frac{1}{K} + \sum_{n=N_1+1}^{N}\frac{1}{K} \tag{7}$$

into Eq. (6), we arrive

$$T(x) = T_0 + \frac{(T_1-T_0)}{1+\frac{N_2 S_2}{N_1 S_1}} = T_0 + (T_1-T_0)x + (T_1-T_0)\left[\frac{1}{\frac{(1-x)}{x}\cdot z + 1} - x\right], \tag{8}$$

where random variables $S_1$, $S_2$, and $Z$ are defined as

$$S_1 = \frac{1}{N_1}\sum_{n=1}^{N_1}\frac{1}{K}, \quad S_2 = \frac{1}{N_2}\sum_{n=N_1+1}^{N}\frac{1}{K}, \tag{9}$$

and

$$z = S_2/S_1. \tag{10}$$

From Eq. (8), the temperature field can be simply written as

$$T(x) = \underbrace{T_0 + (T_1-T_0)x}_{1} + \underbrace{(T_1-T_0)(W-x)}_{2}, \tag{11}$$

where we introduced another random variable $W$ that is a function of random variable $z$,

$$W = \frac{xS_1}{xS_1 + (1-x)S_2} = \frac{1}{1 + \frac{1-x}{x}z}. \tag{12}$$

The temperature field $T$ has two contributions (indicated in Eq. (11) as 1 and 2). Obviously, term 1 represents the homogenized solution by solving the homogeneous heat conduction problem and term 2 represents the stochastic contribution due to the randomness of $K$.

Up to this stage, we do not use any statistical information of the random field $K$. solutions (11) and (12) are applicable to any arbitrary random field $K$. For the purpose of demonstration, thermal conductivity $K$ at any given position $x$ is assumed to follow an independent and identical distribution $f_K$ (i.i.d) with a mean of $\mu_K$ and a standard deviation of $\sigma_K$. Next, we can find the probability distribution function for $\xi = 1/K$,

$$\xi = \frac{1}{K} \sim \frac{f_K(\xi)}{\xi^2}. \tag{13}$$

Clearly, $S_1$ and $S_2$ are the means of $\xi$ in the range of [0, $x$] and [$x$, 1], respectively, and independent of each other. The mean $\mu_\xi$ and variance $\sigma_\xi^2$ of random variable $\xi$ can be found for any arbitrary distribution $f_K$,

$$\mu_\xi = E(\xi) = \int_0^\infty f_K(\xi)/\xi \, d\xi, \tag{14}$$

and

$$\sigma_\xi^2 = E(\xi^2) - E^2(\xi) = \int_0^\infty f_K(\xi) d\xi - \left(\int_0^\infty f_K(\xi)/\xi \, d\xi\right)^2. \tag{15}$$

Two Random variables $S_1$ and $S_2$ are the sum of many independently and identically distributed variable $\xi$ (Eq. (9)). Therefore, we can make use of the central limit theorems (CLT) if both numbers $N_1$ and $N_2$ are sufficiently large. The distributions of variables $S_1$ and $S_2$ are approximately normal for sufficiently large $N_1$ and $N_2$ regardless of the functional form of probability distribution of variable $\xi$,

$$S_1 \to N\left(\mu_\xi, \frac{1}{N_1}\sigma_\xi^2\right) \text{ and } S_2 \to N\left(\mu_\xi, \frac{1}{N_2}\sigma_\xi^2\right), \tag{16}$$

where $\to$ denotes convergence in the probability distribution. Since $S_1$ and $S_2$ represent two normal random variables independent of each other, the random variable $z$ from Eq. (10) should follow the Gaussian ratio distribution and the corresponding probability distribution function is:

$$f_Z(z) = \frac{b(z)c(z)}{a^3(z)} \cdot \frac{1}{\sqrt{2\pi}\sigma_{S1}\sigma_{S2}}\left[2\Phi\left(\frac{b(z)}{a(z)}\right) - 1\right] + \frac{\exp\left(-\left(\mu_{S1}^2/\sigma_{S1}^2 - \mu_{S2}^2/\sigma_{S2}^2\right)/2\right)}{a^2(z)\pi\sigma_{S1}\sigma_{S2}}, \tag{17}$$

where

$$\mu_{S1} = \mu_{S2} = \mu_\xi, \ \sigma_{S1} = \sigma_\xi^2/N_1, \ \sigma_{S2} = \sigma_\xi^2/N_2, \tag{18}$$

$$a(z) = \sqrt{\frac{z^2}{\sigma_{S1}^2} + \frac{1}{\sigma_{S2}^2}}, \ b(z) = \frac{\mu_{S1}}{\sigma_{S1}^2}z + \frac{\mu_{S2}}{\sigma_{S2}^2}, \tag{19}$$

$$c(z) = \exp\left[\frac{1}{2}\frac{b^2(z)}{a^2(z)} - \frac{1}{2}\left(\frac{\mu_{S1}^2}{\sigma_{S1}^2} + \frac{\mu_{S2}^2}{\sigma_{S2}^2}\right)\right] \text{ and } \Phi(x) = \left[erf\left(\frac{\sqrt{2}}{2}x\right) + 1\right]\Big/2. \tag{20}$$

After simplification, we obtain the distribution function for random variable $z = S_2/S_1$,

$$f_Z(z) = \frac{\sqrt{N_1 N_2}}{\pi(N_2 z^2 + N_1)}\left\{\underbrace{e^{-N\mu_\xi^2/(2\sigma_\xi^2)}}_{1} + \underbrace{\sqrt{\frac{\pi}{2}}\alpha \cdot erf\left(\frac{\sqrt{2}}{2}\alpha\right) \cdot e^{-N^*\mu_\xi^2/(2\sigma_\xi^2)}}_{2}\right\}, \tag{21}$$

where two dimensionless number $\alpha$ and $N^*$ are defined as

$$\alpha = \frac{\mu_\xi}{\sigma_\xi}\frac{N_2 z + N_1}{\sqrt{N_2 z^2 + N_1}}, \tag{22}$$

and

$$N^* = \frac{(z-1)^2}{z^2/N_1 + 1/N_2}. \tag{23}$$

Figure 1 presents a plot of variation of $N^*$ with z. It can be shown that $0 \leq N^* \leq N$ and the second term (term 2) in Eq. (21) is dominating over the first term for $z \sim 1$. An approximate PDF around $z \sim 1$ is obtained,

$$F_Z(z) = \frac{\sqrt{N_1 N_2}}{\sqrt{2\pi} \cdot \sqrt{N}} \frac{\mu_\xi}{\sigma_\xi} \cdot e^{-N_1 N_2 (z-1)^2 \mu_\xi^2 / (2N\sigma_\xi^2)} \approx f_Z(z). \tag{24}$$

In order to find the distribution function for random variable W that is a function of z, we make use of the characteristic function $\varphi_W$ of W,

$$\varphi_W(\omega) = E\left(e^{i\omega W}\right) = \int_{-\infty}^{\infty} e^{i\omega W} f_Z(z) dz, \tag{25}$$

where $\omega$ is the frequency. From Eq. (12), we obtain

$$z = \left(\frac{1}{W} - 1\right)\frac{x}{1-x}, \tag{26}$$

and

$$\frac{dz}{dW} = -\frac{1}{W^2} \cdot \frac{x}{1-x}. \tag{27}$$

Substitution of Eqs. (26) and (27) into Eq. (25) leads to the characteristic function

$$\varphi_W(\omega) = \int_{-\infty}^{\infty} e^{i\omega W} \frac{x}{W^2(1-x)} f_Z\left(\left(\frac{1}{W} - 1\right)\frac{x}{1-x}\right) dW, \tag{28}$$

and the distribution function for W (namely $f_W$) is written as

$$f_W(W) = \frac{x}{W^2(1-x)} f_Z\left(\left(\frac{1}{W} - 1\right)\frac{x}{1-x}\right). \tag{29}$$

After we substitute the approximate distribution $f_z(z) \approx F_Z(z)$ for $z$ (Eq. (24)) into $f_W$ (Eq. (29)), we obtain the probability distribution for random variable $W$

$$f_W(W) = \frac{1}{\sqrt{2\pi}\sigma_W} e^{-\frac{(W-\mu_W)^2}{2\sigma_W^2}}, \tag{30}$$

where the mean and standard deviation are

$$\mu_W = x \text{ and } \sigma_W = \frac{\sqrt{x(1-x)}}{\sqrt{N}} \cdot \frac{\sigma_\xi}{\mu_\xi}, \tag{31}$$

with $\mu_\xi$ and $\sigma_\xi$ determined from Eqs. (14) and (15) for any given distribution $f_K$. The distribution of $W$ simply follows a normal distribution with position-dependent mean and variance. The distribution of $W$ is only dependent on $\mu_\xi$ (first order moment) and $\sigma_\xi$ (second order moment), instead of the entire probability distribution function of $\xi$ (namely $f_\xi$). Especially, the variance is decreasing with increasing number of random variables $N$ ($\sigma_W^2 \sim 1/N$), and the largest variance of $T$ is found at the middle point where $x = 1/2$.

The statistical properties of steady-state temperature field $T$ can be eventually found through Eq. (11) for any arbitrary random field of $K$ consisting of i.i.d variables with given distribution $f_K$. Obviously $T$ also follows a normal distribution regardless of the probability distribution form of random field $K$. The mean and standard deviation of T are written as,

$$\mu_T = T_0 + (T_1 - T_0)x \text{ and } \sigma_T = (T_1 - T_0)\sigma_W = (T_1 - T_0)\frac{\sqrt{x(1-x)}}{\sqrt{N}} \cdot \frac{\sigma_\xi}{\mu_\xi}. \tag{32}$$

It is noted that both mean and variance of temperature field are position-dependent, with largest variance at the middle location $x = 0.5$. For random field $K$ consisting of i.i.d variables, all we need to determine the variance of temperature $T$ are the $\mu_\xi$ (first order moment of $f_\xi$) and $\sigma_\xi$ (second order moment $f_\xi$).

## III. Stochastic Analysis for Transient Heat Conduction

In this section, results from our previous study on the homogenization of heat conduction with arbitrary position-dependent thermal conductivity by employing the dispersion relation [13, 17] will be used in order to simplify the stochastic analysis. The main results are first summarized and introduced here. The transient heat conduction equation and the corresponding boundary and initial conditions used here are written as

$$C\frac{dT}{dt} = \frac{d}{dx}\left[K\frac{dT}{dx}\right] \text{ for } x \in [0,1], \tag{33}$$

with Dirichlet boundary conditions

$$T(x=0,t) = 0 \text{ and } T(x=1,t) = 0, \tag{34}$$

and the initial condition

$$T(x,t=0) = T_0, \tag{35}$$

where $C$ is the product of material density and the heat capacity. Both $C$ and $K$ can be random fields. We use the Dirichlet boundary conditions (Eq. (34)) for the purpose of demonstration. The general solution for Eq. (33) can be written as [17]

$$T(x,t) = \underbrace{\tilde{T}}_{1} + \underbrace{\sum_{n=1}^{\infty} L_n(x)\frac{\partial^n \tilde{T}}{\partial x^n}}_{2} \tag{36}$$

By making use of the expansion of dispersion relation. Here $\tilde{T}$ is the homogenized solution by solving the homogenized equation with effective properties (refer to [17]), namely equation

$$\tilde{C}\frac{d\tilde{T}}{dt} = \tilde{K}\frac{d^2\tilde{T}}{dx^2} \text{ for } x \in [0,1] \tag{37}$$

subject to the same boundary and initial conditions (Eqs. (34) and (35)). Effective thermal conductivity $\tilde{K}$ and $\tilde{C}$ are written as

$$\tilde{K} = 1\bigg/\int_0^1 1/K\,dy \text{ and } \tilde{C} = \int_0^1 C\,dy. \tag{38}$$

$L_n(x)$ in Eq. (36) are bounded functions of $x$ satisfying $L_n(0) = L_n(1) = 0$. The first three $L_n(x)$ are given in [17],

$$L_1 = (G_1 - x), \tag{39}$$

$$L_2 = F_2 + x^2/2 - G_1(x + F_2(1) - 1/2), \tag{40}$$

$$L_3 = 2F_4 - G_1\big(F_2 + 2F_4(1) + x/2 - F_2(1)(1+x) - x^2/2 - 1/6\big) - F_2 x - x^3/6, \tag{41}$$

where integral functions $G_1$ and $F_1$, $F_2$, and $F_4$ are defined as

$$G_1(x) = \tilde{K}\int_0^x 1/K\,dy, \tag{42}$$

$$F_1(x) = \int_0^x C_n\,dy\bigg/\tilde{C}_n, \tag{43}$$

$$F_2(x) = \tilde{K}\int_0^x F_1/K\,dy, \tag{44}$$

$$F_4(x) = \tilde{K}\int_0^x F_1 G_1/K\,dy. \tag{45}$$

Homogenized temperature filed $\tilde{T}$ can calculated based on the effective properties $\tilde{K}$ and $\tilde{C}$ for any given set of random variables $\{K_i\}$ and $\{C_i\}$, where $i = 1$ to $N$. On the other hand, the same set of $\{K_i\}$ and $\{C_i\}$ leads to the same effective properties $\tilde{K}$ and $\tilde{C}$ according to eq. (38), but $L_n(x)$ will be different due to different permutations. This stochastic contribution (configurational) through $L_n(x)$ cannot be represented by homogenized solution $\tilde{T}$ because $\tilde{T}$ is same for all configurations as long as the effective properties $\tilde{K}$ and $\tilde{C}$ are the same, while $L_n(x)$ can be different for different configurations (permutations). $L_n(x)$ represents configurational contribution in Eq. (36) (term 2).

Similar to the steady-state solution (Eq. (11)), Eq. (36) expresses the total solution $T$ in terms of the homogenized solution $\tilde{T}$ (term 1 on the right hand side of Eq. (36)) representing the ensemble contribution from the entire random media as a whole, while the configurational contribution from variations within the random media (term 2 of Eq. (36)) is represented by random variables $L_n(x)$. Two contributions are at different scales and are independent of each other.

For the simplest situation where we only consider $K$ as a random field and a constant field $C = 1$ is used for Eq. (33), the following simplifications can be made

$$F_1(x) = x, \ F_2(x) = xG_1 - \int_0^x G_1 dy, \ F_4(x) = \frac{1}{2}xG_1^2 - \frac{1}{2}\int_0^x G_1^2 dy. \tag{46}$$

The expressions for $L_n(x)$ are:

$$L_1 = (G_1 - x), \tag{47}$$

$$L_2 = G_1\int_0^1 G_1 dy - \int_0^x G_1 dy + \frac{1}{2}x^2 - \frac{1}{2}G_1 = L_1\int_0^1 L_1 dy + x\int_0^1 L_1 dy - \int_0^x L_1 dy, \tag{48}$$

and

$$L_3 = G_1 \int_0^1 G_1^2 dy - \int_0^x G_1^2 dy + (x+G_1)\int_0^x G_1 dy$$

$$-G_1(x+1)\int_0^1 G_1 dy + \frac{G_1-x^3}{6} + \frac{xG_1(1-x)}{2}. \tag{49}$$

Expressions for $L_n(x)$ (Eqs. (47)-(49)) satisfy the boundary condition $L_n(0) = L_n(1) = 0$. The total solution converges to the homogenized solution ($T \to \tilde{T}$) as $L_n(x) \to 0$ for a sufficiently large number of random variables $N$ ($N \to \infty$), where $\sigma_{G_1} = \sigma_W \to 0$ and $G_1 \to x$. In order to obtain some insights on the uncertainty associated with the temperature field $T$, we can take the expansion Eq. (36) up to the first order,

$$T(x,t) \approx \tilde{T}(x,t) + L_1 \frac{\partial \tilde{T}}{\partial x}. \tag{50}$$

The homogenized solution $\tilde{T}(x,t)$ is stochastic in nature because it is the solution of effective Equation (37), where $\tilde{C} = 1$ and effective $\tilde{K}$ is a random variable. Similarly, we do not use any statistical information of the random field $K$ up to this stage. Solutions (36)-(45) are applicable to any arbitrary random field $K$. Stochastic properties of effective properties $\tilde{K}$ and functions $L_n(x)$ must be determined first for given random field $K$, followed by the calculation for stochastic properties of temperature field $T$. Again for demonstration, thermal conductivity $K$ at any given position $x$ is assumed to follow an independent and identical distribution $f_K$ (i.i.d) with a mean of $\mu_K$ and a standard deviation of $\sigma_K$.

The statistical properties of $\tilde{K}$ can be found from Eq. (38), where we have

$$\frac{1}{\tilde{K}} = \frac{1}{N}\sum_{i=1}^{N}\frac{1}{K}. \tag{51}$$

Using central limit theorem (CLT) we obtain,

$$1/\widetilde{K} \to N\left(\mu_\xi, \sigma_\xi^2/N\right). \tag{52}$$

By applying the delta method, $\widetilde{K}$ was found to asymptotically converge to the normal distribution

$$\widetilde{K} \to N\left(\frac{1}{\mu_\xi}, \frac{\sigma_\xi^2}{N\mu_\xi^4}\right). \tag{53}$$

By comparing expressions of $G_1$ (Eq. (42) and Eq. (36)) and $W$ (Eq. (12) and Eq. (8)), it is obvious that $G_1 = W$, and therefore the statistical properties for $L_1$ are

$$\mu_{L_1} = 0 \text{ and } \sigma_{L_1} = \sigma_W. \tag{54}$$

The mean and the variance of the solution $T$ can be found by

$$\mu_T = E(T) = \mu_{\widetilde{T}} + E\left(\frac{\partial \widetilde{T}}{\partial x} L_1\right), \tag{55}$$

and

$$\sigma_T^2 = E(T^2) - E(T)^2 = E\left(\widetilde{T}^2\right) - E\left(\widetilde{T}\right)^2 + 2E\left(\widetilde{T}\frac{\partial \widetilde{T}}{\partial x} L_1\right) + E\left(\left(\frac{\partial \widetilde{T}}{\partial x} L_1\right)^2\right). \tag{56}$$

By using the statistical properties of $L_1$ (Eq. (54)), we can arrive the final expressions for the mean $\mu_T$ and variance $\sigma_T^2$,

$$\mu_T = \mu_{\widetilde{T}} \text{ and } \sigma_T^2 = \underbrace{\sigma_{\widetilde{T}}^2}_{1} + \underbrace{\sigma_W^2 \cdot E\left[\left(\partial \widetilde{T}/\partial x\right)^2\right]}_{2}, \tag{57}$$

where $\mu_{\widetilde{T}} = E\left(\widetilde{T}\right)$ and $\sigma_{\widetilde{T}}^2 = E\left(\widetilde{T}^2\right) - E\left(\widetilde{T}\right)^2$ is the mean and variance of the homogenized solution $\widetilde{T}$. Obviously, the variance of solution $T$ ($\sigma_T^2$ in Eq. (57)) includes two contributions, namely $\sigma_{\tilde{T}}^2$ (term 1) from the homogenized solution and term 2 from the configurational contribution of $\sigma_W^2$ and gradient of $\tilde{T}$. Obviously, larger temperature gradient ($\partial \tilde{T}/\partial x$) leads to larger uncertainty in temperature $T$ (larger $\sigma_T^2$).

## IV. Numerical Examples

In this section, numerical examples are provided in order to assess the accuracy of the stochastic analysis proposed in Section III. An initial condition of $T_0 = \sin(\pi x)$ is used. The analytical solution of $\tilde{T}$ can be found as,

$$\tilde{T} = e^{-\pi^2 \tilde{K} t} \sin(\pi x) \tag{58}$$

by solving Eq. (37). It is clear that this particular homogenized solution $\tilde{T}$ follows a lognormal distribution with $\tilde{K}$ following normal distribution (Eq. (53)). The mean and variance of $\tilde{T}$ can be found as:

$$\mu_{\tilde{T}} = E(\tilde{T}) = \exp\left[-\frac{\pi^2 t}{\mu_\xi} + \frac{(\pi^2 t)^2}{2N\mu_\xi^2}\left(\frac{\sigma_\xi}{\mu_\xi}\right)^2\right] \sin(\pi x), \tag{59}$$

$$\sigma_{\tilde{T}}^2 = \left\{\exp\left[\frac{(\pi^2 t)^2}{N\mu_\xi^2}\left(\frac{\sigma_\xi}{\mu_\xi}\right)^2\right] - 1\right\} \exp\left[-\frac{2\pi^2 t}{\mu_\xi} + \frac{(\pi^2 t)^2}{N\mu_\xi^2}\left(\frac{\sigma_\xi}{\mu_\xi}\right)^2\right] \sin^2(\pi x), \tag{60}$$

and

$$E\left[(\partial \tilde{T}/\partial x)^2\right] = \exp\left[-\frac{2\pi^2 t}{\mu_\xi} + \frac{2(\pi^2 t)^2}{N\mu_\xi^2}\left(\frac{\sigma_\xi}{\mu_\xi}\right)^2\right][\pi \cos(\pi x)]^2. \tag{61}$$

We first assume a uniform distribution of $K$ within the range of $[a,b]$. The mean and variance of $\xi = 1/K$ can be found from Eqs. (14) and (15),

$$\mu_\xi = \frac{\ln b - \ln a}{b-a} \text{ and } \sigma_\xi^2 = \frac{1}{ab} - \left(\frac{\ln b - \ln a}{b-a}\right)^2. \tag{62}$$

The mean ($\mu_T$) and variance ($\sigma_T^2$) of solution $T$ can be obtained by substitution of Eqs. (59)-(62) and Eq. (31) into the expression (57).

In order to access the accuracy of the proposed analysis, Monte Carlo simulations were also implemented over a large number of sampling (~1000 repetitive tests are used in this work) and numerical results were used as reference solutions for the purpose of comparison. The range of the uniformly distributed thermal conductivity $K$ is $a$=0.1 and $b$=1 and the total number of random variables $N$ is 100. Figures 2 and 3 present the mean and standard deviation of the total solution $T$ at various time $t$=0.2, 0.4, 0.6, 0.8, 1.0, where the solid lines are the reference solutions from Monte Carlo simulations and the dash lines are the solutions from the proposed stochastic analysis (Eq. (57)). The mean $\mu_T$ is in very good agreement with reference solutions for all time $t$. The standard deviation $\sigma_T$ from Eq. (57) captures the spatial variation (double bell shape) and the agreement with reference solutions becomes better with increasing time.

Figure 4 plots the time variation of standard deviation $\sigma_T$ at two locations $x$=0.2 (represented by thin solid and dash lines) and $x$=0.5 (represented by thick solid and dash lines). Better agreement between two solutions is found for $x$=0.5 and the discrepancy is decreasing with time for $x$=0.2. The spatial variation of variance of random variable $L_1(x)$ (Eqs. (54) and (31) for $\sigma_{L_1}$) is also presented in Fig. 5, where a very good agreement is obtained compared against the reference solution from direct Monte Carlo simulations.

The effect of *N* (total number of random variables) on the accuracy of this analysis is also investigated by running the direct Monte Carlo simulations with *N*=10. Obviously, a larger variance of $L_1$ ($\sigma_{L_1} \propto 1/\sqrt{N}$, see Fig. 8) leads to large discrepancies between this stochastic analysis and MC simulations for the mean (Fig. 6) and standard deviation (Fig. 7) of solution *T*.

**V. Conclusion**

A novel stochastic analysis for steady and transient one-dimensional heat conduction problem is presented based on the homogenization approach. Both steady and transient solutions of *T* are solved analytically for arbitrary random thermal conductivity field consisting of i.i.d variables. Through the expansion of *T* around the homogenized solution $\tilde{T}$, it was shown that uncertainty for the stochastic temperature field can be separated into the ensemble contribution (through the effective thermal conductivity $\tilde{K}$, and hence the homogenized temperature $\tilde{T}$) and the configurational contribution (through random variables $L_n(x)$). Expressions of $L_1(x)$ and its mean and variance are presented for random field with i.i.d variables. Results are in reasonable agreement with the direct Monte Carlo simulations. For random filed consisting of correlated variables, the stochastic analysis presented here are still applicable (steady-state solution (11) and transient solution (50)) in general. The Karhunen-Loeve decomposition may be needed to identify the statistical properties of temperature field *T* based on solutions (11) and (50). This will be included in our future work.

Figure 1. The variation of $N^*$ with random variable $Z$.

Figure 2. Variation of mean solution $\mu_T$ at various time $t=0.2, 0.4, 0.6, 0.8, 1.0$ for $N=100$. Solid lines represent the reference solutions from Monte Carlo simulations. Dash lines represent solutions from this stochastic analysis (Eq. (57)).

Figure 3. Variation of standard deviation $\sigma_T$ at various time $t=0.2, 0.4, 0.6, 0.8, 1.0$ for $N=100$. Solid lines represent reference solutions from Monte Carlo simulations. Dash lines represent solutions from this stochastic analysis (Eq. (57)).

Figure 4. Variation of standard deviation $\sigma_T$ with time at $x=0.2$ and $0.5$. Solid lines represent reference solutions from Monte Carlo simulations. Dash lines represent solutions from this stochastic analysis (Eq. (57)).

Figure 5. Spatial variation of standard deviation of variable L$_1$ ($\sigma_{L_1} = \sigma_W$) with $x$ for $N=100$. Solid lines represent reference solutions from Monte Carlo simulations. Dash lines represent solutions from this stochastic analysis (Eq. (31)).

Figure 6. Variation of mean solution $\mu_T$ at various time $t=0.2, 0.4, 0.6, 0.8, 1.0$ for $N=10$. Solid lines represent the reference solutions from Monte Carlo simulations. Dash lines represent solutions from this stochastic analysis (Eq. (57)).

Figure 7. Variation of standard deviation $\sigma_T$ at various time $t$=0.2, 0.4, 0.6, 0.8, 1.0 for $N$=10. Solid lines represent reference solutions from Monte Carlo simulations. Dash lines represent solutions from this stochastic analysis (Eq. (57)).

Figure 8. Spatial variation of standard deviation of variable $L_1$ ($\sigma_{L_1} = \sigma_W$) with $x$ for $N$=10. Solid lines represent reference solutions from Monte Carlo simulations. Dash lines represent solutions from this stochastic analysis (Eq. (31)).

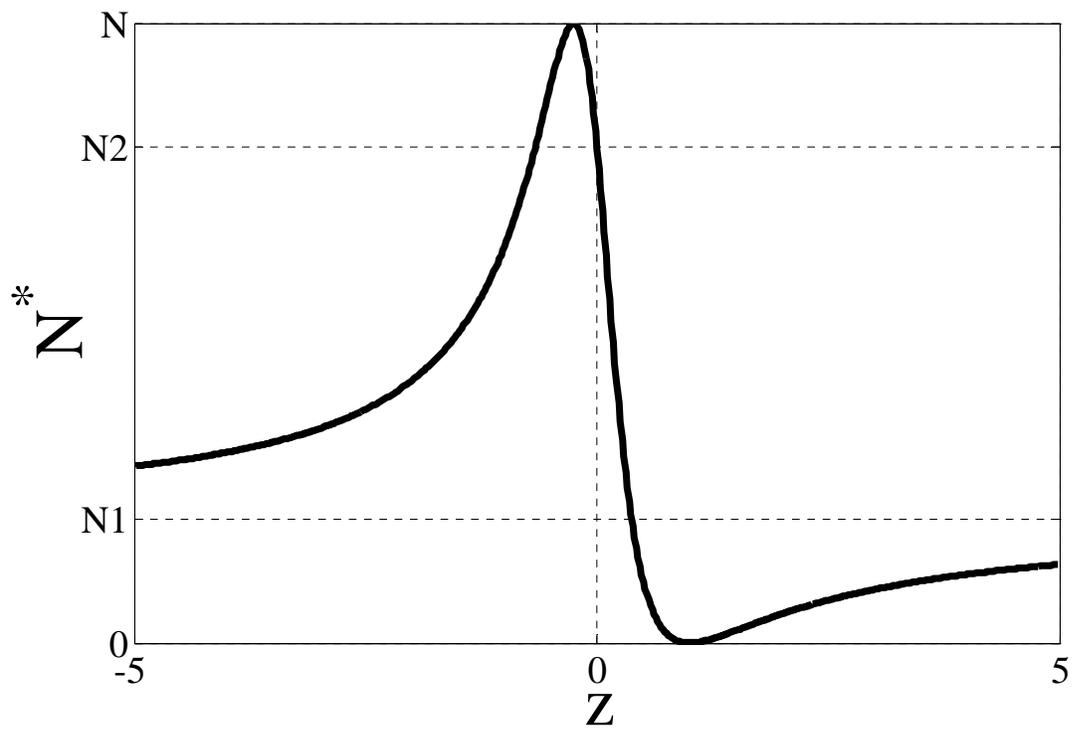

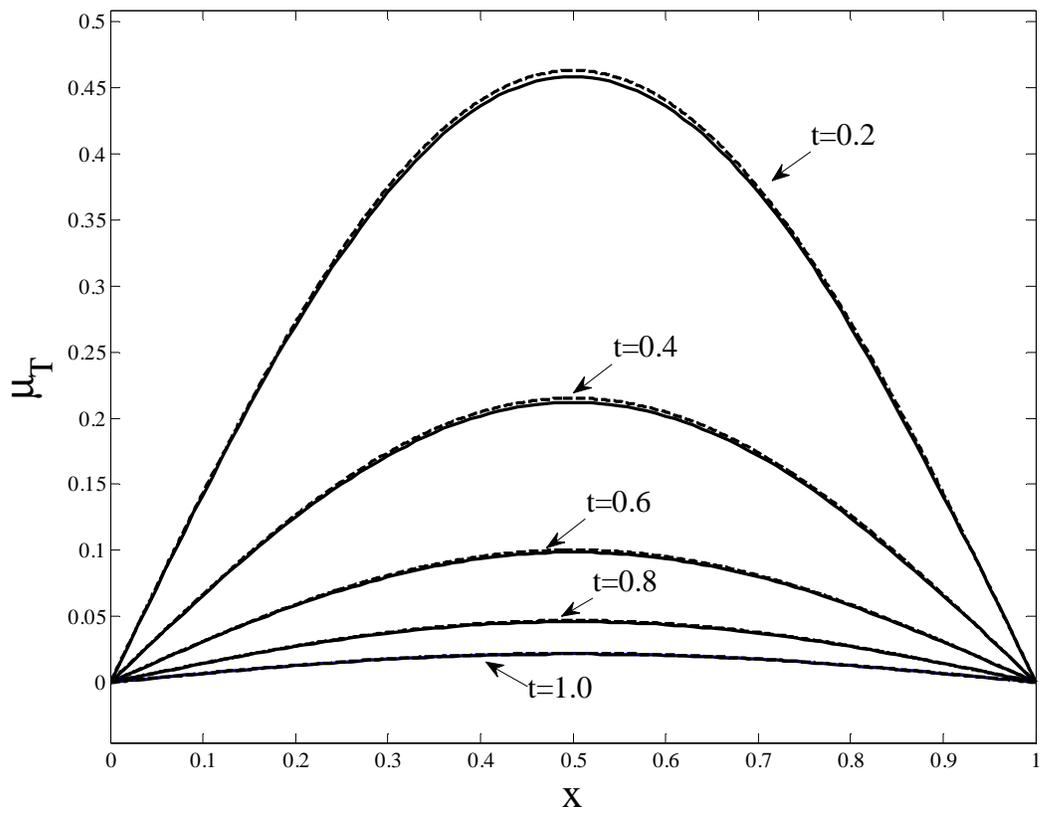

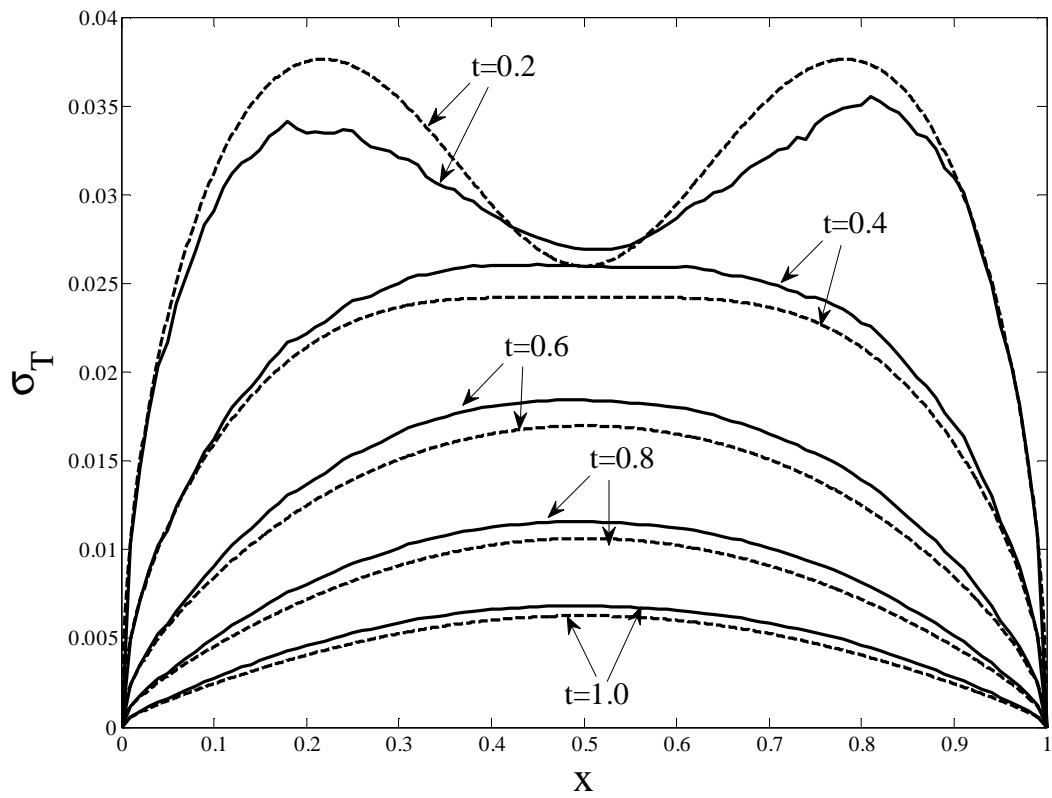

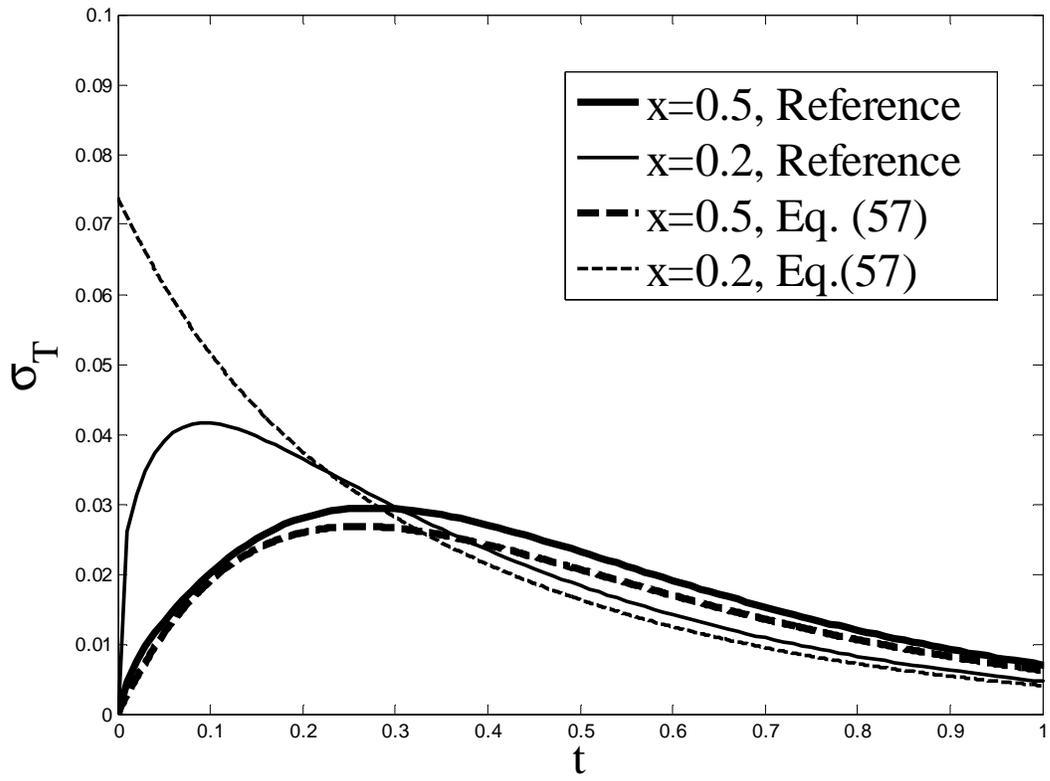

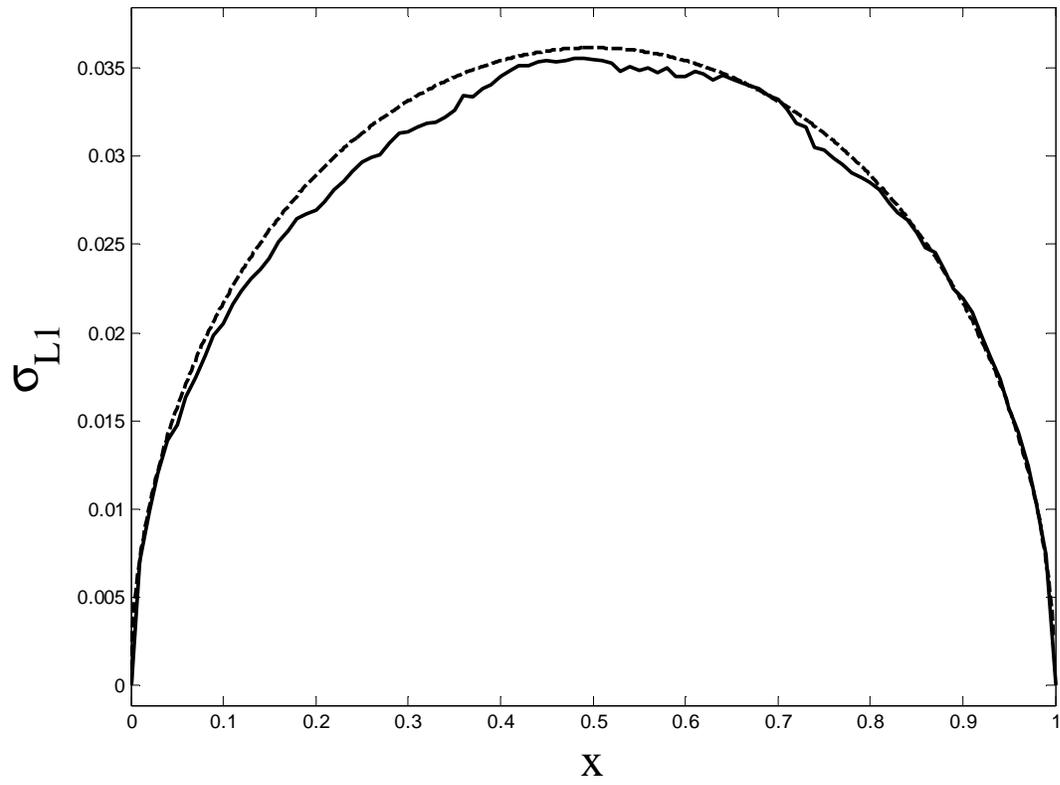

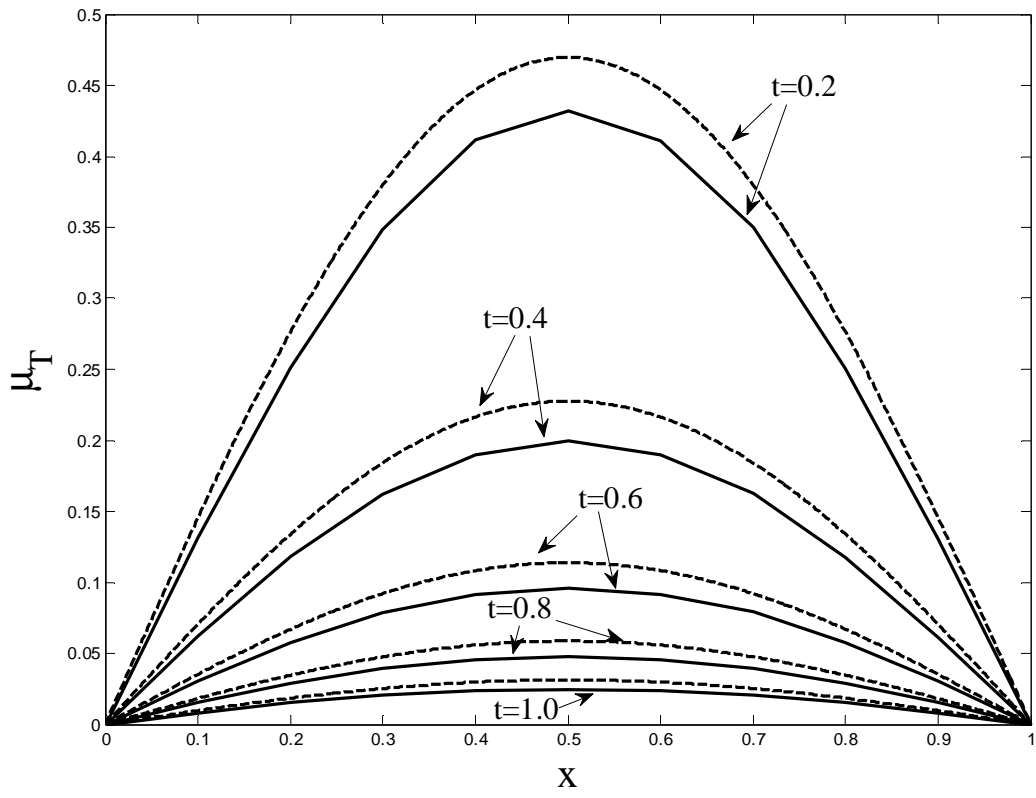

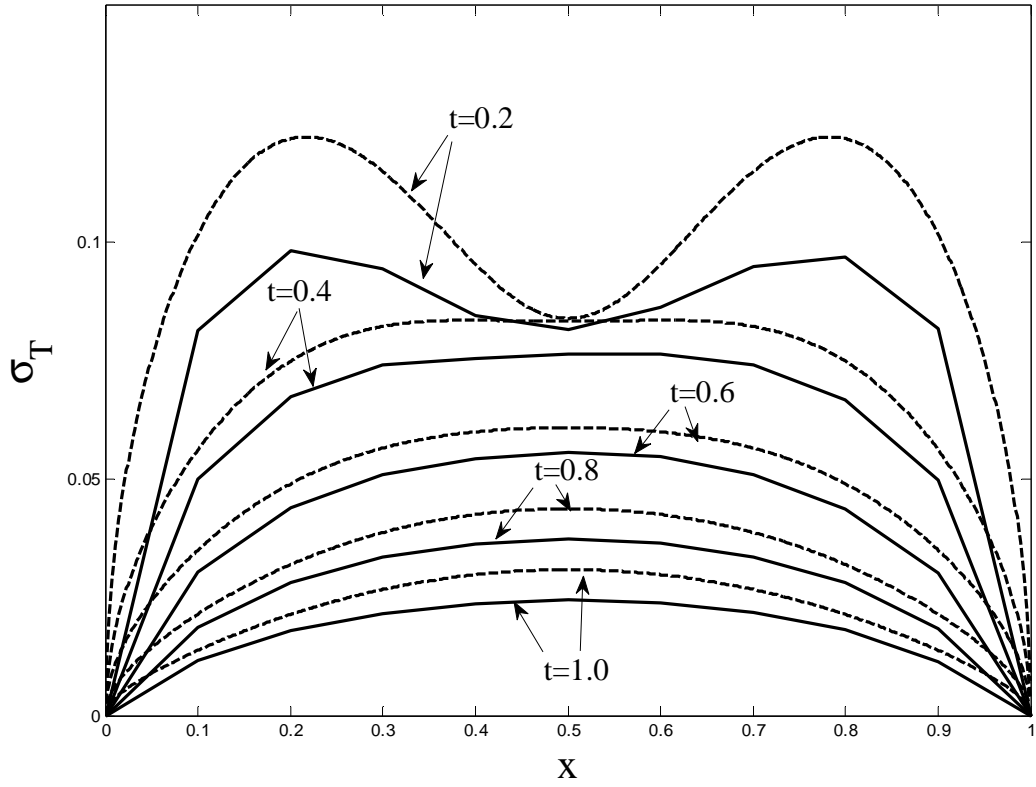

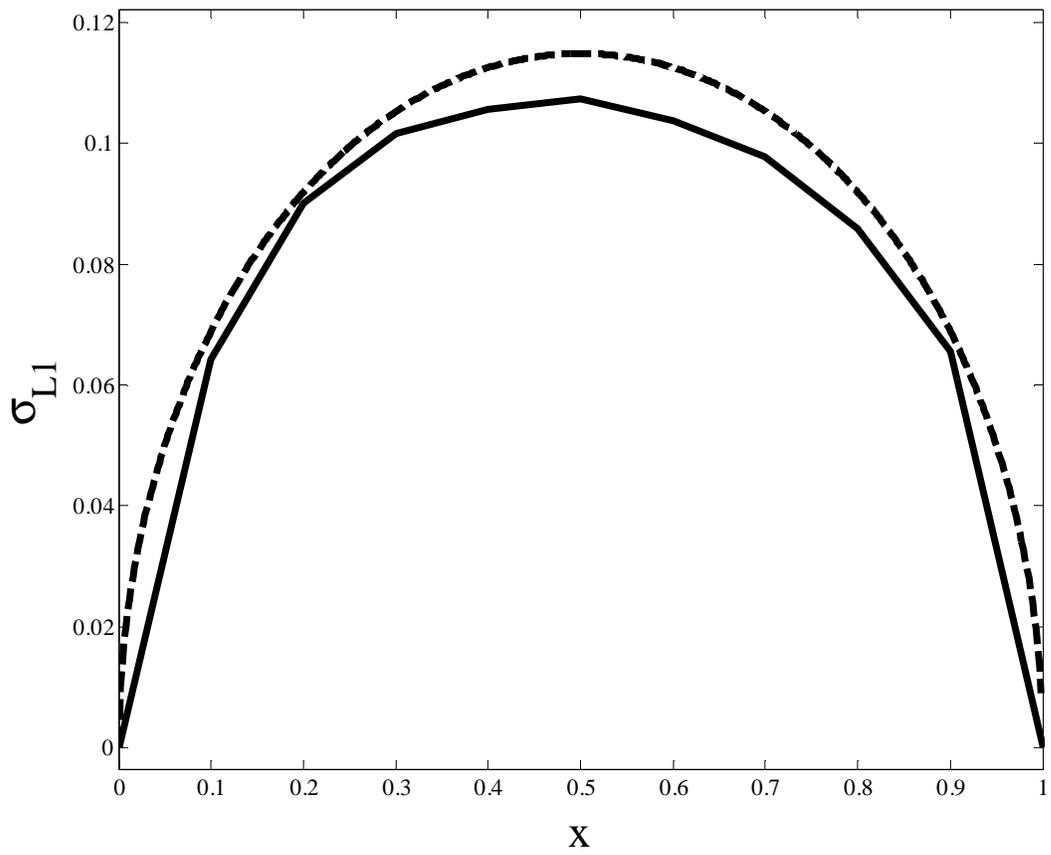